# MASSIVE STARS IN THE SDSS-IV/APOGEE SURVEY. I- OB STARS

A. Roman-Lopes,[1] C. Román-Zúñiga,[2] Mauricio Tapia,[3] Drew Chojnowski,[4] Y. Gómez Maqueo Chew,[5]
D. A. García-Hernández,[6,7] Jura Borissova,[8,9] Dante Minniti,[9,10,11] Kevin R. Covey,[12] Penélope Longa-Peña,[13]
J. G. Fernandez-Trincado,[14,15] Olga Zamora,[6,7] and Christian Nitschelm[13]

[1]*Department of Physics & Astronomy - Universidad de La Serena - Av. Juan Cisternas, 1200 North, La Serena, Chile*
[2]*Instituto de Astronomía, Universidad Nacional Autónoma de México, Unidad Académica en Ensenada, Km 103 Carr. Tijuana-Ensenada, Ensenada 22860, México*
[3]*Instituto de Astronomía, Universidad Nacional Autónoma de México, Ensenada, Apartado Postal 877, Ensenada 22860, México*
[4]*Department of Astronomy, New Mexico State University, Las Cruces, NM 88003, USA*
[5]*Instituto de Astronomía, Universidad Nacional Autónoma de México, Ciudad Universitaria, 04510, Ciudad de México, México*
[6]*Instituto de Astrofísica de Canarias, E-38205 La Laguna, Tenerife, Spain*
[7]*Departamento de Astrofísica, Universidad de La Laguna (ULL), E-38206 La Laguna, Tenerife, Spain*
[8]*Instituto de Fisica y Astronomia, Facultad de Ciencias, Universidad de Valparaiso, Valparaiso, Chile.*
[9]*Millennium Institute of Astrophysics, MAS.*
[10]*Departamento de Ciencias Físicas, Facultad de Ciencias Exactas, Universidad Andrés Bello, Av. Fernandez Concha 700, Las Condes, Santiago, Chile.*
[11]*Vatican Observatory, V00120 Vatican City State, Italy.*
[12]*Department of Physics & Astronomy, Western Washington University, 516 High Street, Bellingham, WA 98225-9164, USA*
[13]*Unidad de Astronomía, Universidad de Antofagasta, Avenida Angamos 601, Antofagasta 1270300, Chile*
[14]*Departamento de Astronomía, Universidad de Concepción, Casilla 160-C, Concepción, Chile*
[15]*Institut Utinam, CNRS UMR6213, Univ. Bourgogne Franche-Comté, OSU THETA , Observatoire de Besançon, BP 1615, 25010 Besançon Cedex, France*

## ABSTRACT

In this work we make use of DR14 APOGEE spectroscopic data to study a sample of 92 known OB stars. We developed a near-infrared semi-empirical spectral classification method that was successfully used in case of four new exemplars, previously classified as later B-type stars. Our results agree well with those determined independently from ECHELLE optical spectra, being in line with the spectral types derived from the "canonical" MK blue optical system. This confirms that the APOGEE spectrograph can also be used as a powerful tool in surveys aiming to unveil and study large number of moderately and highly obscured OB stars still hidden in the Galaxy.

## 1. INTRODUCTION

The study of hot, massive stars, from O type dwarfs to their evolved counterparts (blue super-giants, hyper-giants and/or Wolf-Rayet stars) is of utmost importance to understand star formation and further evolution processes. As supernova event progenitors, they define the chemical histories of most stellar populations, while their rapid and violent evolution cycle results in a significant alteration of their local environment through intense feedback. The absence or presence of massive stars in nascent stellar groups defines the difference between a rapid or quiescent removal of the gaseous component (e.g. Román-Zúñiga et al. (2015); Rivera-Galvez et al. (2015). Furthermore, the winds of massive stars affect their surroundings to the point of determining even the formation of planetary systems in neighboring stars (Balog et al. 2008; Walsh, Millar & Nomura 2013). Massive stars dominate the luminosity distribution in clusters and thus are beacons of recent star formation activity from nearby to very distant regions. In fact, in many cases they are the basic tracers of the local star formation efficiency at distinct scales. Although intrinsically very luminous, when immersed deep in dense molecular clouds, they can be extremely faint at optical wavelengths due to high obscuration ($A_V = 10-20$ mag). In such cases, obtaining near-infrared spectra is the best option for their classification and analysis, as they are in principle easy to detect in this range (Roman-Lopes, Barba & Morrell 2011; Roman-Lopes 2012, 2013a,b; Roman-Lopes, Franco & Sanmartin 2016).

Many questions about O-stars still remain unanswered, among them: what are the physical properties and processes in the regions where they form (Liu et al. 2015; He et al. 2016)?, what separates them from those where only low mass stars are produced?, What is the current number of O stars in the Milky Way?, what is their environmental distribution, e.g., clusters versus distributed population versus runaways? These questions are possible motivators for two lines of action: one is to expand the search for unclassified O stars, and another one is to augment our understanding of their infrared spectra in order to know their earliest phases of evolution by reaching their most obscured sites (Przybilla 2010).

As O stars are the rarest among all spectral types, the spectroscopic confirmation of a single exemplar is an achievement on itself. In fact, to date less than 1000 O-type stars are confirmed and classified as such (Maíz-Apellániz et al. 2013). Moreover, the atmospheres of O-stars are the least stable among all spectral types, a fact that complicates atmospheric modeling further, for example, accounting for the simultaneous action of heavy mass loss, rotation, and non-LTE effects when modeling their atmospheres (e.g. Nieva & Przybilla 2010).

Determining spectral classification of hot massive stars in the infrared is advantageous in terms of dust extinction, but has a few caveats. As discussed by Blum et al. (1997), photospheric absorption features are scarce and generally weak compared to other lines used to classify late-type stars. Moreover, several insterstellar and circumstellar absorption features can be present, particularly in the K band, (Blum et al. 1997; Meyer et al. 1998). In this sense, the H-band near 1.6 $\mu$m is an attractive alternative.

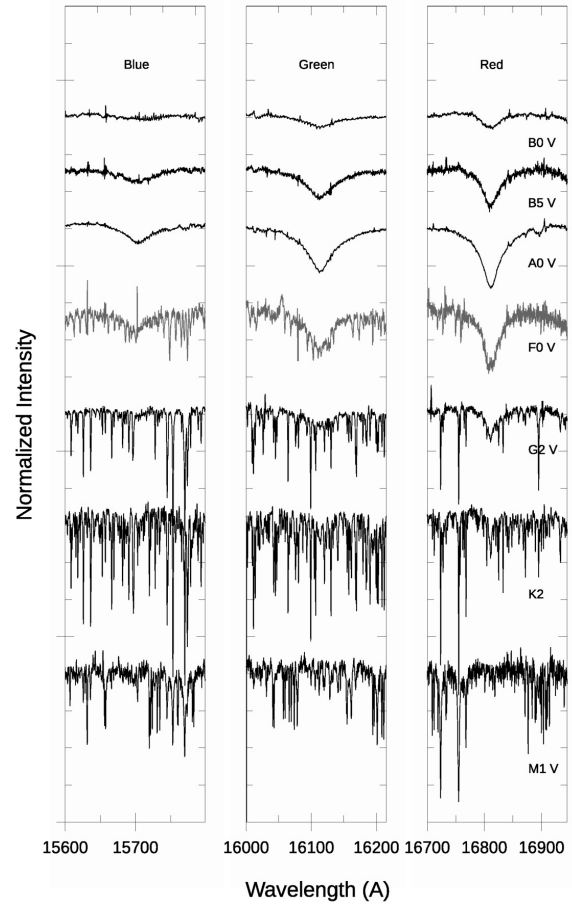

**Figure 1.** Blue, green and red portions of the APOGEE spectra of known B0 (2M00225332+6214290), B5 (2M02500186+5906467), A0 (2M00171874+6249332), F0 (2M0156746+3729086), G2 (2M01435156-1833554), K2 (2M00220599+8540038), and M1 (2M00182256+4401222) main sequence stars.

In this paper we study a set of spectra of O- and early-B stars taken with the APOGEE spectrograph based on Data Release 14 (hereafter DR14 – Abolfathi et al. 2017, submitted). Our primary goal is to demonstrate the viability of using APOGEE spectra to identify O and early B-type stars and also to provide reliable and accurate estimates of their spectral types and luminosity classes. In Section 2 we briefly describe the APOGEE-2 spectroscopic survey, and how it was obtained; in Section 3 we detail the identification and classification of our OB star sample, translating the numerical results into the canonical MK optical spectral system. In Sections 4 and 5 we discuss and conclude with our main results.

## 2. HOT STARS IN SDSS-IV

### 2.1. *The APOGEE Survey*

The Apache Point Observatory Galactic Evolution Experiment (APOGEE Majewski et al. 2017) was developed as one of the core surveys of the third version of the Sloan Digital Sky Survey - SDSS-III - (Eisenstein et al. 2011; Gunn et al. 2006; Houtzman 2015). It aims at obtaining nearly 130 thousand spectra of red giant stars from all Galactic components in order to study the history, structure, kinematics and chemical evolution of the Milky Way Galaxy. Currently, the second phase of the program (APOGEE-2 - Blanton et al. (2017)) extends the sample to the Southern hemisphere sky from the Las Campanas Observatory, and thus contemplates an unprecedented All-Sky spectral database of over 300 thousand sources, that includes partial coverage of high extinction regions towards the Galactic center and Bulge, and even the Magellanic Clouds (Zasowski et al. 2017).

Each of the two APOGEE (Wilson et al. 2010) instruments (one for the north and one for the south) is a 300-fiber spectrograph conceived to work in the near-infrared (NIR) on the blue portion (~15000Å-17000Å) of the H-band. With a resolving power R~22500 it works over three non-overlapping detectors covering the ranges 15145Å-15810Å (blue), 15860Å-16430Å (green), and 16480Å-16950Å (red). Each APOGEE observation or visit is a 1-hour block integration that covers one field of 1.0-1.5° in radius. The fiber bundles are normally spread among the field to obtain three kind of targets: (i) *science* targets (230 fibers), (ii) *sky* spectra that are used to remove the air-glow effects on the data (35 fibers), and (iii) a *telluric* sample (35 fibers) composed of observations of bright blue stars whose spectra are used to remove the absorption effects produced by the Earth's atmosphere on the data. For details regarding APOGEE data reduction procedure, see Nidever et al. (2015); Garcia Perez et al. (2016).

### 2.2. *APOGEE spectra of known OB-stars*

The main atmospheric contaminants in the 1.51-1.70 $\mu$m APOGEE wavelength range are OH air-glow emission as well as carbon dioxide, water and methane absorption bands. The survey strategy has been to include, in each plate and visit, a set of 30-35 blue stars to characterize telluric absorption. As a result, the SDSS-III survey obtained a large number of high-resolution H-band spectra of stars that served as telluric standards (for details on the APOGEE survey target selection, see Zasowski et al. 2013). Among this main sample of blue objects with APOGEE data, we selected a sub-sample of high signal-to-noise (S/N) H-band spectra (usually S/N > 250-300) composed of 92 known OB-stars (hereafter named APOGEE-OB sample), almost half of them being (41) O type stars. For this sample, we obtained spectral types and luminosity classes directly from the literature. In Table 1 we list these objects along with their identifications, H-band photometry, spectral types and luminosity classes. For this table we list, when available, the uncertainties from the corresponding paper (see list of references in Table 1); otherwise, we adopted an uncertainty value of one subclass, which is conservative for typical optical spectral type classification methods.

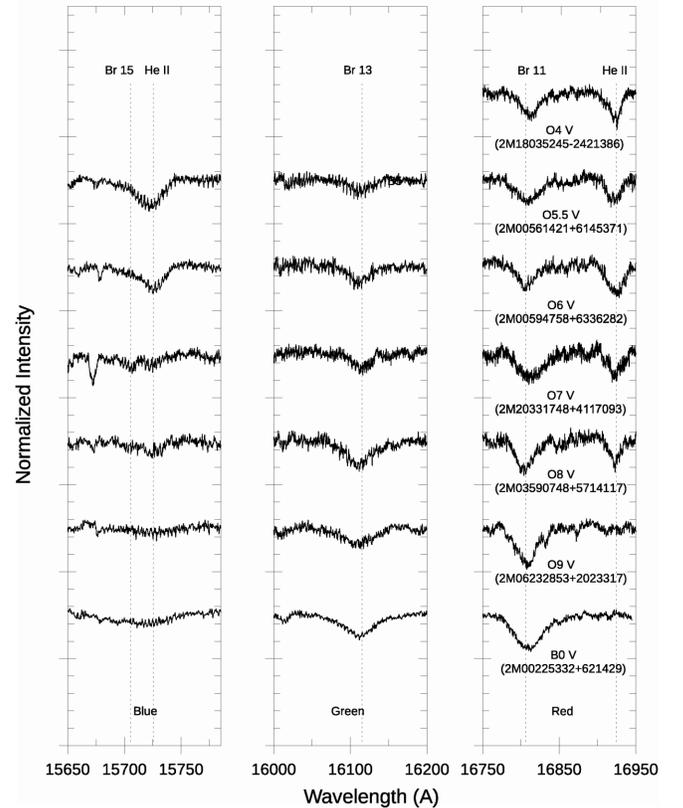

**Figure 2.** APOGEE spectra of known main sequence OB-stars for stars of spectral types ranging from O4 V (with only red detector useful data from DR14) to B0 V. In this figure it is shown the portions of the APOGEE spectral range that contain the hydrogen and He lines of interest to our work. In the vertical scale, the separation between major ticks corresponds to 10% of the normalized continuum scale.

An important characteristic of the APOGEE-OB sample is the lack of strong "metallic" absorption lines, when compared with those detected in the spectra of F-, G-, K and M-stars. To illustrate this, in Fig.1 we present blue, green and red APOGEE spectra of main sequence stars of spectral types ranging from M1 V to B0 V. As can be seen there, the later the spectral type is, the more numerous and intense the lines produced by elements heavier than hydrogen (H) and helium (He) are. Also, we see that the hotter the star is, the less intense are the associated lines of the Brackett series, a feature typically used to identify and separate A and mid-B stars from the OB earlier types. The earliest spectral types can be easily identified and separated from the later ones due to the lack (at least to the APOGEE sample sensitivity) of metallic lines of heavier elements like Na, Ca, Mg, Fe, etc. On the other hand, it is well known that A-type stars in general present the strongest hydrogen line transitions. Indeed, from this figure we can see that the Br11, Br13 and Br15 lines are much more intense than those observed in the spectrum of the B5-star, which also has lines stronger than those observed in the APOGEE spectrum of the B0-star, which presents the weakest absorption lines.

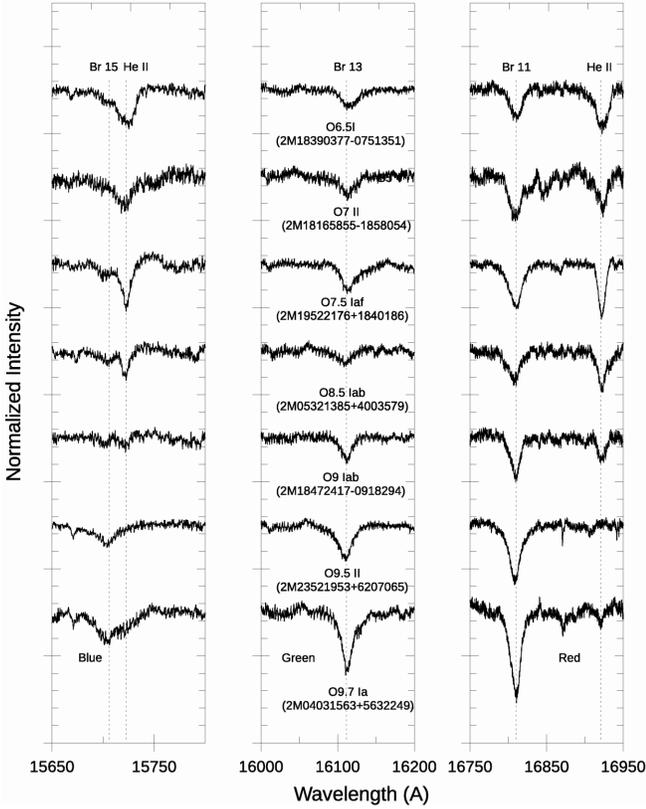

**Figure 3.** Apogee spectra of APOGEE-OB star sample corresponding to mid-O (O6.5) to late-O (O9.7) type stars, and luminosity class corresponding to super-giant stars. In the vertical scale, the separation between major ticks corresponds to 10% of the normalized continuum scale.

In Figures 2 and 3, we present some representative spectra of the APOGEE-OB star sample corresponding to early-O (O4) to early-B (B1) type stars, and luminosity classes ranging from dwarfs (Figure 2) to super-giants (Figure 3). The main features of interest for our work are those useful for the identification and (when possible) classification of OB-stars: (i) The hydrogen lines of the Brackett series $\lambda16811$ (11-4), $\lambda16113$ (13-4), and $\lambda15705$ (15-4), and (ii) two HeII lines, $\lambda15723$ (7-13) and $\lambda16923$ (7-12). The latter vanishes for spectral types later than O9-O9.5 in dwarfs, sub-giants and giant classes. For the dwarfs, we notice that the HeII $\lambda15723$ (7-13) and $\lambda16923$ (7-12) lines vanishes for spectral types later than O8-O9, with the hydrogen lines of the Brackett series becoming stronger from the earlier to the later types. Also noticeable is the presence of very strong HeII $\lambda15723$ (7-13) absorption in the case of the earliest O-types. For the super-giants, it is possible to see how the relative intensities of the HeII lines, as well as, the Brackett lines strength change from the later to the earliest O-types. On the other hand, from the comparison of the hydrogen line profiles present in the spectra, it is possible to see that the hydrogen and helium lines are in general (much) broader for stars of class V than for those of class I to class II. Also, from both figures it is possible to deduce that the relative intensities of the HeII lines, as well as, the Brackett lines strength, change from the later to the earliest O-types. Finally, from the comparison of the hydrogen line profiles seen in the spectra in Figures 2 and 3, it is possible to conclude that the H and He lines are in general (much) broader for stars of class V than for those of class I to class II.

### 2.3. *Main Spectral features in the APOGEE-OB star sample*
#### 2.3.1. *Line parameters measurements*

Table 2 contains the results from the measurements of the equivalent width (EW), and full width half maximum (**FWHM**) of the hydrogen (HI Br11 and Br13) and helium (HeII 7-12 and 7-13) lines observed in the spectra of the APOGEE-OB sample. We followed standard procedures to obtain these measurements, that we briefly describe as follows. In order to obtain the line-profile parameters, we used several spectroscopic analysis routines available in the IRAF[1] SPLOT package. For each spectral line, we performed model fittings using both Voigt and pure Gaussian profiles, with the associated continuum computed from regions far from the observed line wings (in general several tens of angstroms from the estimated line centers), with the final uncertainty in the EW and **FWHM** values ranging from about 10% for spectra with the highest S/N (usually above 500) to 15-20% for the spectra presenting lower S/N values in the range 250-500.

---

[1] http://iraf.noao.edu/

Table 1. OB Stars in the APOGEE Survey

| Star | APOGEE ID | ID (Literature) | H mag (2MASS) | SpType | Lum. Class | Qual. | Ref.[a] |
|---|---|---|---|---|---|---|---|
| 1 | 2M00022269+6254032 | LS I +62 59 | 9.34 | B0.5 | III | | (1) |
| 2 | 2M00175039+5903271 | HD 1334 | 7.81 | B2.5 | V | | (30) |
| 3 | 2M00200554+6203587 | HD 1544 | 7.79 | B0 | III | | (19) |
| 4 | 2M00225332+6214290 | HD 1810 | 7.99 | B0 | V | n | (41) |
| 5 | 2M00481254+6259249 | ALS 6351 | 10.22 | O7.5 | V | z | (3) |

[a] List of known APOGEE O- and early B-stars used in this work. Column (1) contains the internal ID; column 2 the APOGEE ID; column 3 the ID from the literature; column 4 the 2MASS H-band magnitudes; column 5 the spectral type taken form the literature; column 6 the corresponding luminosity class; column 7 the associated qualifier; and column 8 the reference from which the spectral types and luminosity classes were taken. References: (1) Negueruela & Marco (2003), (2) Cakirli et al. (2014) (3) Maíz-Apellániz et al. (2016) (4) Garmany & Vacca (1991) (5) Barbier (1968) (6) Hiltner (1956) (7) Sota et al. (2014) (8) Hoag (1965) (9) Mathys (1989) (10) Lesh & Aizenman (1973), (11) Morgan, Code & Whitford (1955) (12) Russeil, Adami & Georgelin (2007) (13) Sota et al. (2011) (14) Houk & Swift (1999) (15) Voroshilov et al. (1985) (16) Garrison, Hiltner & Schild (1977) (17) Davis (1977) (18) Walborn (1971) (19) Baraya (1979) (20) Hiltner & Iriarte (1955) (21) Radoslavova (1989) (22) Straizyz et al. (2014) (23) Kiminki et al. (2007) (24) Garrison & Kormendy (1976) (25) Crampton, Georgelin & Georgelin (1978) (26) Massey, Johnson & Degioia-Eastwood (1995) (27) Houk (1982) (28) Grenier et al. (1999) (29) Houk & Smith-Moore (1988) (31) Lehmann (2011) (32) Jaschek & Jaschek (1980) (33) Guetter (1968) (34) Cowley (1972) (35)Massey, Degioia-Eastwood & Waterhouse (2001) (36) Lesh (1968) Georgelin (1967) (38) Suad et al. (2016) (39) Abt & Corbally (2000) (40) Chavarria-K et al. (1988). The entire version of the table is available in the electronic version of the article.

We notice that the He II 7-12 $\lambda16923$ line (when present) is sometimes seen close to the edge of the APOGEE spectral band (at about 16954 Å), which at this level may complicate the estimation of the continuum (at the red side of the line) by the algorithms of the IRAF tasks. In such cases, we proceeded to measure the line parameters carefully "by hand", using the IRAF task tools in iterative mode. When not even this procedure was possible, the results were simply discarded. Fortunately, only for a few cases (corresponding to early O-stars with no He II 7-12 values quoted in Table 2), we were unable to properly measure that spectral line.

Table 2. Line Properties for OB Stars in APOGEE[a]

| Star | H I (Br11) EW (Å) | H I (Br11) FWHM (Å) | H I (Br13) EW (Å) | H I (Br13) FWHM (Å) | He II (7-12) EW (Å) | He II (7-12) FWHM (Å) | He II (7-13) EW (Å) | He II (7-13) FWHM (Å) |
|---|---|---|---|---|---|---|---|---|
| 1 | 2.3 | 22.7 | 2.1 | 29.0 | | | | |
| 2 | 3.6 | 32.7 | 3.2 | 55.3 | | | | |
| 3 | 1.1 | 20.1 | 1.0 | 19.1 | | | | |
| 4 | 1.66 | 36.8 | 1.1 | 45.5 | | | | |
| 5 | 1.0 | 41.5 | 0.33 | 30.4 | 0.35 | 16.7 | 0.26 | 19.8 |

[a] Equivalent width and full width half maximum measurements for the the H I (Br11 and Br13) and He II (7-12 and 7-13) spectral lines detected in the APOGEE2-OB sample. Associated uncertainties on the quoted values are estimated to vary from 10 to 15 percent. The entire table is available in the electronic version of the paper.

2.4. *Spectral types, luminosity classes and the observed line parameters dependences: O-stars*

In this section, we present the results obtained from our analysis of the line parameter measurements for O-type stars observed with APOGEE.

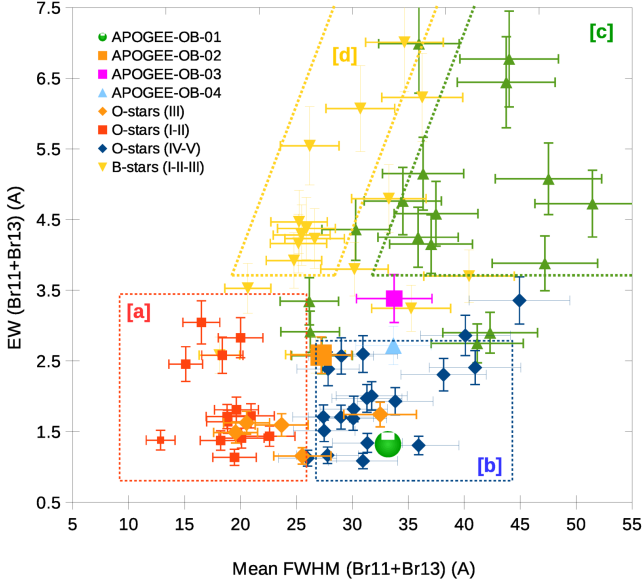

**Figure 4.** The EW[Br11+Br13] versus mean FWHM[Br11+Br13] diagram made from data in Table 2. Region [a] - O-stars of classes I-II (red squares). Region [b] - O-stars of classes III-IV-V (orange and blue diamonds). Region [c] - B-stars of classes IV-V (green triangles). Region [d] - B-stars of classes I-II-III (yellow triangles).

2.4.1. *Spectral types and luminosity classes of OB stars based on the APOGEE line parameters*

In general terms, the spectra of stars earlier than O9 display, besides the HI Brackett lines, the 7-12 and 7-13 He II lines, making it possible to estimate luminosity classes and spectral types from the associated EW and **FWHM** measured values. Figure 4 shows the EW[Br11+Br13] versus mean FWHM[Br11+Br13] diagram in which the O-stars of luminosity classes I-II and III are represented, respectively, by red and orange squares, while the O-stars of luminosity classes IV-V are shown as blue circles. On the other hand, B-stars are represented by yellow (class I-II-III) and green (class IV-V) triangles. We can see that there is a clear separation between B- and O-type stars, the former spread in a region of the diagram well above the O-star sample, as these occupy the bottom part of the diagram.

In fact, we find that four groups of stars occupy well-defined areas of the EW[Br11+Br13] versus mean FWHM[Br11+Br13] plot. These are delimited by colored dotted lines in Fig. 4. *Region (a)* (red) is occupied by O-type stars of luminosity classes I-II-III with the extreme super-giants concentrated in its upper part, *Region (b)* (blue) by O-type stars mainly of the sub-giant and dwarf classes, *Region (c)* (green) by B-type stars of luminosity classes IV-V, and *Region (d)* (yellow) is dominated by giant to super-giant B-type stars. From the diagram, we infer that: (i) There is a clear separation between O-stars of classes I-II from classes IV-V, with the class III members sharing most of the space of the first group, with a few remaining exemplars found among the last. (ii) The few O- and B-type stars seen in the interfaces between regions (a)-(d), and regions (b)-(c) belong to the O9.5 to B0-B1 spectral types.

2.4.2. *HI Brackett lines*

In Figure 5, we show the EW vs. Spectral Type (SpType) diagrams made from the EW-Br11 and EW-Br13 values listed in Table 2. From the observed distributions of sources on both diagrams, it is possible to separate the stars in three distinct groups:

(i) A first set is formed exclusively by O stars ("O-stars only"), corresponding to sources with APOGEE spectra where the H Brackett lines have equivalent width values satisfying the criteria EW(Br11) < 1.0 Å and EW(Br13) < 0.75 Å. We notice that in the case of the earliest O-type stars (like the two O5V seen in Figure 4(b)) the measured Br13 EW values probably go beyond the mentioned limit. Nevertheless, it is possible to discriminate early O-type stars from the B-type ones based also on the presence of the HeII 7-12 and 7-13 line transitions.

(ii) A second group ("[O + B0-B1] stars") presenting EW measured values in the range 1.0 < EW(Br11) < 2.1 Å, and 0.75 < EW(Br13) < 1.5 Å, which is formed by OB stars of spectral types in the range O7-O8 to B0-B1.

(iii) Finally, a third group presenting EW values satisfying the criteria EW(Br11) > 2.0 Å and EW(Br13) > 1.5 Å. This last group contains only B-type stars ("B-stars-only").

Interestingly, in case of the third group, it is possible to see that the strength of the hydrogen Br11 and Br13 lines diminish monotonically from later to earlier spectral types. Indeed, the observed linear relation is quite good, mainly for B stars earlier than $\approx$ B3-B3.5. In Figure 6 we show the EW[Br11+Br13] versus SpType diagram for giants to supergiants stars (left), together with that for sub-giant and dwarf stars (right). Both diagrams were made using H line parameters taken from Table 2. Considering the observed trend in both diagrams, it is possible to conclude that there is a well-defined linear relation between EW[Br11+Br13] and SpType in the case of early- to mid-B stars. Based on linear fittings to the data in each diagram, we have the following empirical linear relations:

(a) Giants and super-giants (B0-B3)

$$SpType\ [class\ I, II, III] = 0.56 \times EW[Br11+Br13] + 8.99 \quad (1)$$

(b) sub-giants and dwarfs (B0-B5)

$$SpType\ [class\ IV, V] = 0.64 \times EW[Br11+Br13] + 8.47 \quad (2)$$

The numerical results from this equations relate to the spectral types in accordance with the correspondence shown in Figure 5, i.e. 10 corresponds to B0, 11 to B1,..., 15 to B5.

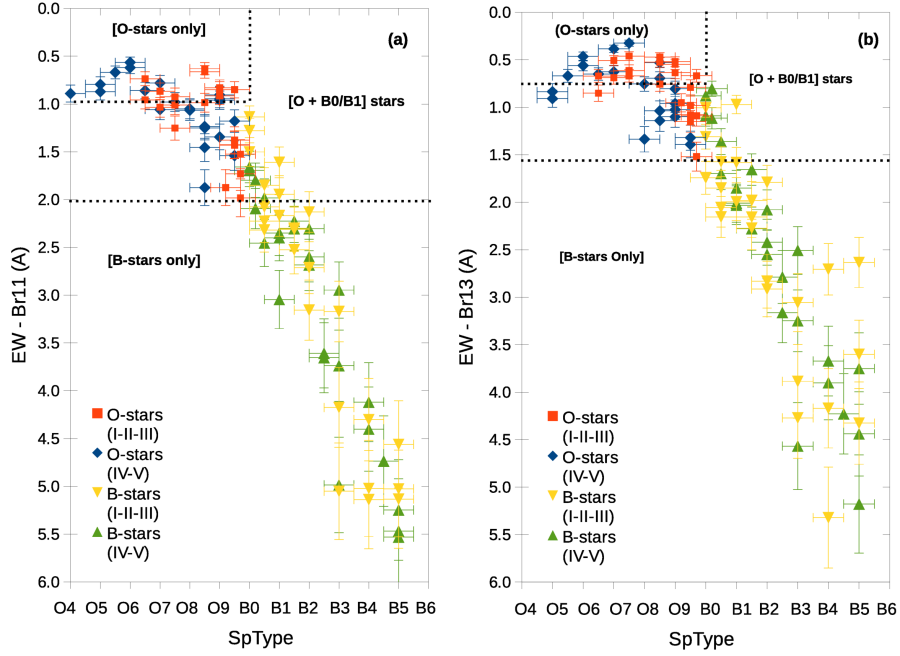

**Figure 5.** The APOGEE [Br11-Br13] EW versus SpType diagrams made from the corresponding values in Table 2. O and B class I - III stars are indicated by red squares and yellow triangles, respectively. O IV-V and B IV-V stars are indicated, respectively, by the blue circles and green triangles.

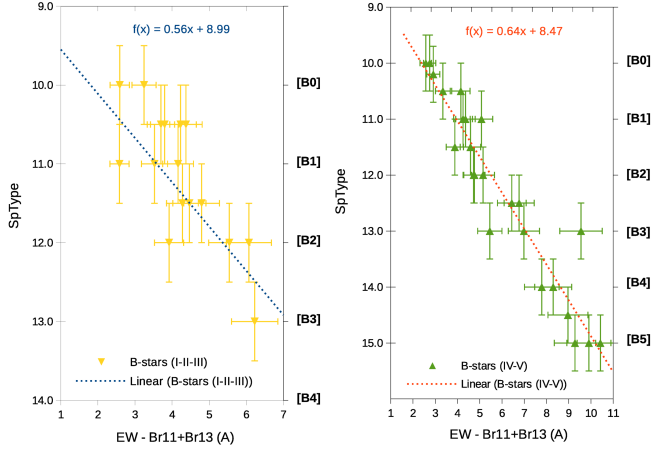

**Figure 6.** The SpType versus EW[Br11+Br13] diagrams for B-type giants and super-giants (left) and B-type sub-giants and dwarfs (right). Both diagrams were made using hydrogen line parameters taken from Table 2. The trend of the points in both diagrams indicates a quite good linear relation between EW[Br11+Br13] and SpType in the case of early- to mid-B stars.

### 2.4.3. *He II 7-12 and He II 7-13 lines*

The two HeII lines seen in the APOGEE spectra of O-type stars are generated by the 7-12 and 7-13 transitions, and from the line measurements presented in Table 2 we can see that the former vanishes in O-stars belonging to luminosity classes IV-V, and spectral types later than O9-O9.5 (e.g. sources # 18, 29 and 91 - O9V and sources # 25, 78 and 85 - O9.5IV-V). The only notable exception is source # 20, an O9V star that does show a weak He II 7-12 line at $\lambda16923$, and no He II absorption line at $\lambda15723$, that disappear for all O-stars of spectral types later than O8.5. On the other hand, the cases of giants and super-giants are quite different, and from Table 2 (as well as from the spectra shown in Figure 3), we note that both HeII lines are present, sometimes even in case of the O9.5-O9.7 later types.

In Figure 7 the SpType versus EW[He II] diagrams for O stars (HeII (7-12 (left) and 7-13 (right)), constructed from the EW line measurements of Table 2 are shown. In these diagrams, stars of luminosity classes I-II-III, and luminosity classes IV-V (from now group 1 and group 2) are represented, respectively, by red squares and blue circles. We can see that O-type stars earlier than O8 appear separate in both diagrams, with the sources splitting into two distinct sequences. The first formed by O5 to O9 dwarfs and sub-giants, follows steeper sequences than those of group 2 sources. Linear regressions on the data for groups 1 and 2 yield the following empirical linear relations between spectral types and He II equivalent line widths:

(a) Dwarfs and Sub-giants

$$SpType\ [He\ II\ (7-12)] = -5.05 \times EW[7-12] + 9.32 \quad (3)$$

$$SpType\ [He\ II\ (7-13)] = -6.40 \times EW[7-13] + 9.10 \quad (4)$$

(b) Giants and super-giants

$$SpType\ [He\ II\ (7-12)] = -3.50 \times EW[7-12] + 9.59 \quad (5)$$

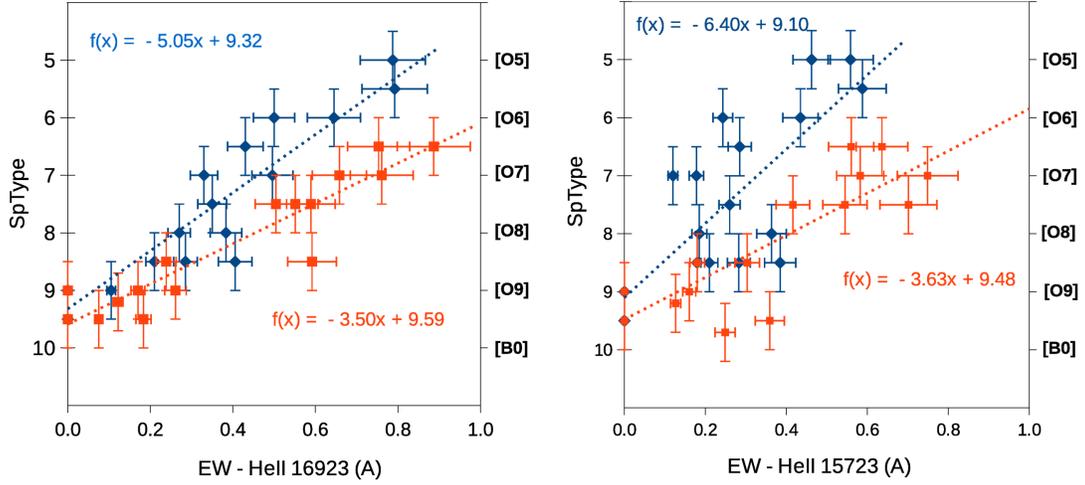

**Figure 7.** The SpType versus He II EW diagrams for O stars, using the data for He II lines (7-12 (left) and 7-13 (right)) present in Table 2. Class I-II-III (group 1) and Class IV-V (group 2) stars are represented respectively by red squares and blue circles.

$$SpType\ [He\ II\ (7-13)] = -3.63 \times EW[7-13] + 9.48 \quad (6)$$

Again, the numbers relate to the spectral types following the correspondence shown in Figure 7, e.g. 5 corresponds to O5, 6 to O6,..., 9 to O9, and 10 to B0.

## 3. IDENTIFICATION AND CLASSIFICATION OF OB STAR CANDIDATES USING APOGEE DATA

In order to further demonstrate the plausibility of using APOGEE spectra not only to identify O-B stars, but also to provide reliable estimates of their spectral types and luminosity classes, we applied the procedures described in Section 2 to a number of new OB stars with spectra taken within the APOGEE survey.

### 3.1. *New O stars in the APOGEE DR14 sample*

From visual inspection of APOGEE DR14 stellar spectra, we selected four examples, cataloged in the literature as B0-B1 stars, which turn out to pertain to spectral types earlier than those previously reported. In Table 3 we list their identifications, H-band magnitudes, spectral types and luminosity classes (if any), all taken from the literature.

Figure 8 displays the "blue", "green" and "red" sections of the APOGEE spectra (only those portions of the spectra centered on the main spectral features used for the identification/classification of OB stars are shown) of the four selected stars. The He II 12-7 and 13-7 lines are seen in absorption in the spectra of stars APOGEE-OB-01 and APOGEE-OB-02, while the spectra of stars APOGEE-OB-03 and APOGEE-OB-04 show no evident He II 7-12 lines, although their Br11 and Br13 EW values are compatible with stars of spectral type between O9 and B1. For this reason, we have included them in the analysis.

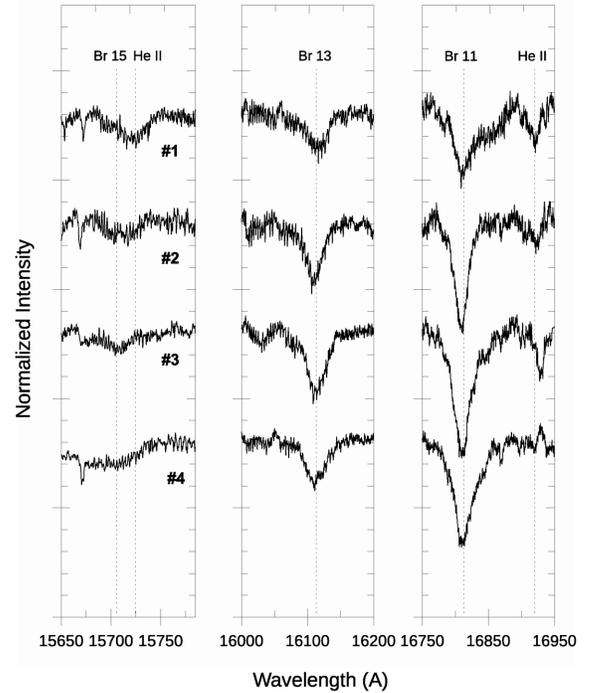

**Figure 8.** The blue, green and red portions of APOGEE-OB-01 to APOGEE-OB-04 (see Table 3) spectra (centered on the main spectral features used in the identification of O stars) of four candidate O stars. In Table 4 we present the values of the line parameters measured from their APOGEE spectra. In the vertical scale, the separation between major ticks corresponds to 10% of the normalized continuum scale.

In Table 4 we present the measured values of the spectral line parameters for stars APOGEE-OB-01 to 04, from the corresponding APOGEE spectra. In Figure 4 these four stars are plotted with special symbols. Their positions in the diagram indicates that only source APOGEE-OB-01 has a high probability of being a class IV-V star. Star APOGEE-OB-02 falls just on the border of regions (a) and (b), so the best we can do in this case is to assign its luminosity class as III-V. Star APOGEE-OB-03 is found outside region (b) in the interface with region (c), precluding the assignment of a luminosity class. Finally, considering that star APOGEE-OB-04 is located inside region (b), very close to several class IV-V template stars, we assign a class IV-V to it.

Now that we have estimates of the luminosity classes of stars APOGEE-OB-01 to 04, it is possible to derive their spectral types by using their hydrogen and helium EW values into Equations 1-6, which in turn are used to transform the EW values from equations 1-6 into spectral type in the canonical way, with EW values of [(5.0-5.5),(5.5-6.0),...,(9.0-9.5)] corresponding to [(O5-O5.5),(O5.5-O6),...,(O9-O9.5)], as well as values [(10-10.5), (10.5-11),...,(14.5-15)] corresponding to [(B0-B0.5),(B0.5-B1),...,(B4.5-B5)]. In the case of star APOGEE-OB-01, we get from equations 3 and 4, SpType [He II]= 7.75, which corresponds to O7.5-O8 in the MK system. For star APOGEE-OB-02, of luminosity class III-V, the mean obtained from equations 3-6, results in SpType [He II]= 8.55, corresponding to O8.5-O9.

Table 3. OB Star candidates in APOGEE DR14 data

| APOGEE-OB | 2MASS | star ID | H (2MASS) | SpType | Lum. Class | Ref.[a] |
|---|---|---|---|---|---|---|
| 01 | 2M00582309+5939149 | BD+58 142 | 9.05 | B0 | — | Kopylov (1953) |
| 02 | 2M05273340+3427018 | LS V +34 21 | 8.13 | B0 | IV | Jose et al. (2008) |
| 03 | 2M06582818-0301252 | HD 51756 | 7.48 | B1 | V | Abt (2008) |
| 04 | 2M06152540+1901459 | HD 253983 | 8.72 | B1 | V | Abt & Corbally (2000) |

[a]Reference for spectral type and luminosity class

As mentioned before, the majority of the O9-O9.5 class IV-V stars (with the exception of star #20) does not show the He II 7-12 or He II 7-13 lines normally seen in the spectra of earlier O-type stars. Thus for stars APOGEE-OB-03 and 04, despite having no detectable He lines in their APOGEE spectra, a late O-type cannot be ruled out, as their Br11 and Br13 EW values satisfy the criteria EW(Br11) < 2.0 *and* EW(Br13) < 1.6. In these two cases, therefore, it is also possible to estimate lower limits to their spectral types by using Equation 3 and 5, setting He II 7-12=0.

We found it impossible to reliably estimate the luminosity class of star APOGEE-OB-03, but from the mean values from equations 3 and 5, an implied lower limit for its spectral type is SpType EW[HeII]=9.2, which corresponds to O9-O9.5. On the other hand, in order to estimate the upper limit for the spectral type of this star, we make use of its EW[Br11+Br13] value into equations 1-2, obtaining SpType EW[Br11+Br13]=10.75, equivalent to B0.5-B1. Considering these two results and the above corresponding sequences, we then assigned spectral type O9-B1 to star APOGEE-OB-3.

Finally, for star APOGEE-OB-04, EW[Br11] = 2.09 and EW[Br13] = 0.7, therefore EW[Br11 + Br13] = 2.79. This value is lower than the limit EW[Br11 + Br13] = 3.6 of the "[O+B0/B1] stars" locus discussed in Section 2.4.1. However, its EW[Br11] value is a bit higher than that observed for the bulk of "O- and B-stars" in Figure 4. In this case there are two possibilities: (i) by assuming that it is of a very early class IV-V B-star, using Equation (2) we obtain [EW(Br11+Br13)]=10.2, corresponding to a MK type B0-B0.5, or (ii) where it is a very late O-star, then from Equation 3 (as it seems to be class IV-V) we get EW[7-12]=0, from which we compute SpType EW[He II]=9.3, that corresponds to spectral type MK O9-O9.5.

**Table 4.** Line parameters for the OB Star candidates and estimated spectral types using APOGEE DR14 data

| APOGEE OB | EW(A) [Br11] | FWHM(A) [Br11] | EW(A) [Br13] | FWHM(A) [Br13] | EW(A) [He II 7-12] | FWHM(A) [He II 7-12] | EW(A) [He II 7-13] | FWHM(A) [He II 7-13] | SpType + Lum. Class |
|---|---|---|---|---|---|---|---|---|---|
| 01 | 0.68 (0.08) | 26.6 (2.8) | 0.64 (0.08) | 39.8 (4.2) | 0.30 (0.1) | 10.2 (1.1) | 0.21 (0.03) | 14.7 (1.5) | O7.5-O8 IV-V |
| 02 | 1.69 (0.19) | 24.3 (2.6) | 0.89 (0.09) | 30.1 (3.3) | 0.18 (0.05) | 9.1 (1.0) | 0.17 (0.03) | 15.2 (1.7) | O8.5-O9 III-V |
| 03 | 1.87 (0.25) | 31.6 (3.3) | 1.51 (0.20) | 35.9 (4.0) | — | — | — | — | O9-B1 |
| 04 | 2.09 (0.22) | 31.7 (3.3) | 0.70 (0.08) | 35.6 (3.7) | — | — | — | — | B0-B0.5 IV-V |

### 3.2. *Optical spectroscopy for the APOGEE-OB-01 to APOGEE-OB-04 stars*

In this section, we test the accuracy and reliability of the method of estimating spectral types/luminosity classes from APOGEE spectra as described in Section 3.1 by comparing those with the corresponding classification from independent spectra with similar resolution, but in the blue optical regime.

#### 3.2.1. *Optical spectra: San Pedro Martir Echelle observations and ESO Feros archive data*

Optical spectra for sources APOGEE-OB-01, 02 and 04 were obtained with the SPM Echelle spectrograph and a Marconi-4 2048×2048 CCD detector mounted on the 2.1m Telescope of the Observatorio Astronómico Nacional on Sierra San Pedro Mártir in Baja California, México. We used a slit width of 200 $\mu$m (about 2.75″) and a cross dispersing grating of 300 lines/mm with a 3600 Å cutoff filter, together with a 13″ mask separating contiguous orders. The resultant spectral resolution is R≈18000 within orders 30-61, corresponding to the 3600< $\lambda$/Å < 7000 range. For star #1 we obtained three 15-min integrations, while for stars APOGEE-OB-02 and 04, two 15-min integrations were sufficient. Each observation was preceded and followed by a 150-s Th-Ar-Ne lamp exposure to allow for wavelength calibration at each possible flexure.

**Table 5.** Line parameters for the OB Star candidates in the blue optical range

| APOGEE-OB | He $\lambda$4026 | Si $\lambda$4089 | He $\lambda$4144 | He $\lambda$4200 | He $\lambda$4387 | He $\lambda$4471 | He $\lambda$4542 | Si $\lambda$4552 | He $\lambda$4686 | He $\lambda$4713 |
|---|---|---|---|---|---|---|---|---|---|---|
| 01 | 0.58 (0.07) | — | 0.07 (0.01) | 0.36 (0.05) | 0.25 (0.03) | 0.67 (0.08) | 0.42 (0.06) | — | 0.65 (0.07) | 0.11 (0.02) |
| 02 | 0.62 (0.08) | 0.38 (0.05) | 0.19 (0.03) | 0.44 (0.05) | 0.39 (0.05) | 0.73 (0.08) | 0.34 (0.04) | — | 0.48 (0.06) | 0.26 (0.04) |
| 03 | 0.72 (0.09) | 0.24 (0.03) | 0.39 (0.05) | 0.11 (0.02) | 0.47 (0.05) | 0.91 (0.11) | 0.13 (0.02) | 0.11 (0.02) | 0.34 (0.04) | 0.21 (0.03) |
| 04 | 1.08 (0.12) | 0.15 (0.02) | 0.47 (0.06) | — | 0.61 (0.07) | 1.17 (0.13) | < 0.05 | 0.10 (0.02) | 0.12 (0.02) | 0.19 (0.03) |

The reduction of the SPM Echelle data was done using standard techniques by using the IRAF ONEDSPEC, TWODSPEC, and APEXTRACT packages. The one-dimensional spectra of the science targets were extracted from the two-dimensional frames by summing pixels in the data range and subtracting off the background value for each order column, which was measured as the median of contiguous non-target pixels. Cosmic rays and other anomalous signal detections were suppressed from each of the extracted spectra by removing pixels that deviate 5-$\sigma$ or more from the mean within a 100 pixel wide box that steps through the spectrum. The bad pixels were replaced through a linear interpolation of the removed data range, and the wavelength calibration was performed using the ThArNe lamp spectra. The final optical Echelle spectra for stars APOGEE-OB-01, 02 and 04 were then normalized through the fitting of the continuum emission in the associated wavelength ranges.

In case of star APOGEE-OB-03, we used ESO archive data. The observation was made in 2003-01-16 in the framework of the ESO program 70.D-0110(A) (PI Poretti), using the FEROS Echelle spectrograph (R ≈ 48000) coupled to the ESO-1.5m telescope. For the reduction of this spectrogram, we followed the same procedure as in the case of the San Pedro Martir Echelle spectrograms described above, normalizing the final spectrum by the fitted continuum in the appropriate wavelength ranges.

In Figure 9 we present the normalized San Pedro Martir ECHELLE and ESO-FEROS blue optical spectra (4000A-4700A) of the four stars. They are shown in decreasing order of effective temperature (from the top to the bottom). The He

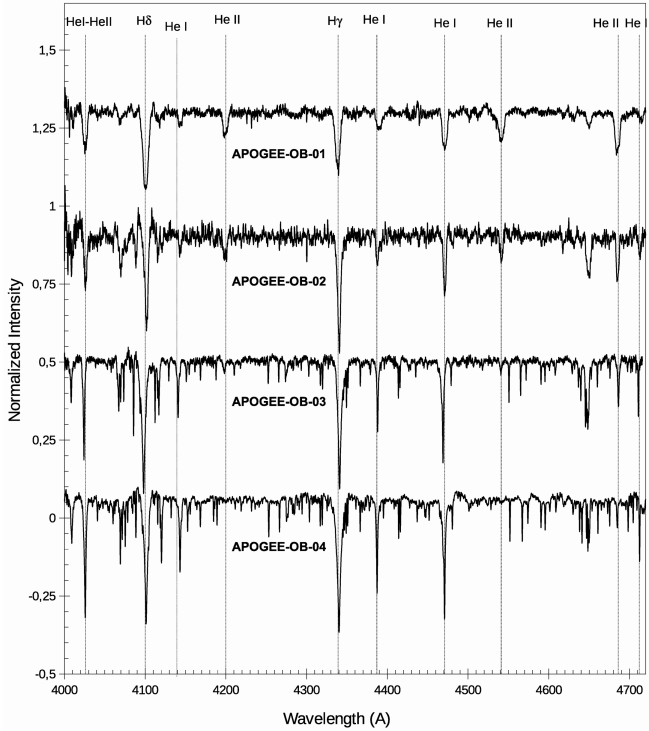

**Figure 9.** Blue optical spectra taken for stars APOGEE-OB-01 to 04, with some of the H I, He I and He II lines found in the associate wavelength range, being indicated.

II lines $\lambda 4200$ and $\lambda 4542$ clearly seen in the first three spectra, while they vanish in case of star APOGEE-OB-04. On the other hand, the He I line $\lambda 4471$ presents the opposite behavior, becoming stronger from the earlier to the later types.

We measured the equivalent widths (EW) of the spectral lines in these blue spectra and estimated independent spectral types for the four stars. In Table 5 we present the EW values for the following absorption lines: (i) eight He lines corresponding to the He I-II $\lambda 4026$, He I $\lambda 4121$, He II $\lambda 4200$, He I $\lambda 4387$, He I $\lambda 4471$, He II $\lambda 4542$, He II $\lambda 4686$, He II $\lambda 4713$ transitions, and (ii) three Si lines corresponding to the Si IV $\lambda 4089$, Si IV $\lambda 4116$ and Si III $\lambda 4552$ transitions. In Table 6, we list the computed EW ratio values used in the optical spectral classification procedure (Conti & Alschuler (1971); Conti & Frost (1977); Mathys (1988); Walborn & Fitzpatrick (1990); Sota et al. (2011)) of the APOGEE-OB-01 to APOGEE-OB-04 stars.

### 3.2.2. *APOGEE-OB-01*

As can be seen in Figure 9, APOGEE-OB-01 (BD+58 142) is the earliest-type star of the four. In the past, based on photografic magnitudes it was classified as a B0 star by Kopylov (1953). Despite being a bright (V=9.71) and very blue (U-B = -0.62) northern source (Mermilliod 2006), to date it did not receive any further spectroscopic studies. Its optical spectrum presents very broad and strong He II $\lambda\lambda$ 4542-4686 lines. The relative strengths of the $\lambda\lambda$ 4471-4542 He lines suggests a spectral type probably later than O7 (for which the two lines should appear with the same intensities). On the other hand, from the $\lambda\lambda$ 4144-4200 He lines, it is possible to estimate a spectral type earlier than O9 (for which the two mentioned lines should appear with the same intensities). From the EW ratios [4542/][4387], [4200]/[4144], [4089]/[4026] shown in Table 6, and the criteria of Sota et al. (2011) (their table 4), we conclude that the spectral type of this star is O8 and, based on its very strong absorption He II 4686 line, its luminosity class is V.

### 3.2.3. *APOGEE-OB-02*

APOGEE-OB-02 (LS V +34 21) was previously classified as a B0 star by Jose et al. (2008), who performed an extense UBVI CCD photometric study of the young open cluster Stock 8. As in the case of APOGEE-OB-01, the ratio of the He II $\lambda\lambda$ 4542-4686 lines indicates a spectral type later than O7. On the other hand, from the relative intensities of the $\lambda\lambda$ 4144-4200 He lines, it is safe to state that it has an spectral type earlier than O9. From the EW ratios [4542/][4387], [4200]/[4144], [4089]/[4026] shown in Table 6, and the criteria presented in Sota et al. (2011) we assign an O8.5 spectral type. The He lines in the optical spectrum of star #2 are narrower than those in the star #1 and, considering that its He II $\lambda 4686$ line appears less intense (or strong), we conclude that its luminosity class is probably III (Sota et al. 2011).

**Table 6.** Line ratios and classification for the OB Star candidates in the blue optical range

| APOGEE-OB | [4542]/[4387] | [4200]/[4144] | [4552]/[4542] | [4089]/[4026][a] | [4686]/[4713][a] | HeII 4686[b] | SpType[c] | Lum. class[d,e] |
|---|---|---|---|---|---|---|---|---|
| 01 | 1.7 | 5.0 | 0 | N/A | N/A | Very strong | O8 | v |

*Table 6 continued*

**Table 6** (*continued*)

| APOGEE-OB | [4542]/[4387] | [4200]/[4144] | [4552]/[4542] | [4089]/[4026][a] | [4686]/[4713][a] | HeII 4686[b] | SpType[c] | Lum. class[d,e] |
|---|---|---|---|---|---|---|---|---|
| 02 | 0.9 | 2.4 | 0 | N/A | N/A | Strong | O8.5 | III |
| 03 | 0.3 | 0.3 | 0.8 | 0.3 | 1.62 | N/A | O9.7 | III |
| 04 | 0.1 | 0 | 2.1 | 0.1 | 0.63 | N/A | B0 | V |

[a] Luminosity criteria only for O9-O9.7 stars - Sota et al. (2011)

[b] Strength of the HeII 4686 line - Luminosity criteria for O8-O8.5 stars - Sota et al. (2011)

[c] Accordingly with criteria in Table 4 of Sota et al. (2011)

[d] Based on criteria from Tables 5 and 6 of Sota et al. (2011)

[e] For B-stars we used the criteria described in Walborn & Fitzpatrick (1990); Didelon (1982)

### 3.2.4. *APOGEE-OB-03*

APOGEE-OB-03 (HD 51756) was previously classified (based on photoeletric photometry) as a B0.5 IV star by Morgan, Code & Whitford (1955). Later, based on low resolution optical spectra it was re-classified as a B1 V star by Abt (2008).

From the ESO-FEROS optical spectrum of APOGEE-OB-03 shown in Figure 9, we can see that it looks quite different from those of stars APOGEE-OB-01 and APOGEE-OB-02, with very intense and narrow λ4026, λ4144 and λ4471 He I lines. From the computed [4542]/[4387], [4200]/[4144], [4552],[4542], [4089]/[4026] and [4686]/[4713] ratio values in Table 6, and the criteria presented by Sota et al. (2011) (their Tables 4 and 6) we conclude that APOGEE-OB-03 is an O9.7 III star.

### 3.2.5. *APOGEE-OB-04*

The last optical spectrum in Figure 9 is that for APOGEE-OB-04 (HD 253983). Based on UBV CCD photometry, it was previously classified as a B1V star by Abt & Corbally (2000). Its San Pedro Martir Echelle spectrum is quite similar to that for star APOGEE-OB-03, with the exception that the He II λλ4200-4542 lines are absent. Based on the criteria defined in Table 6 of Sota et al. (2011), together with its very small [4542]/[4387] and [4200]/[4144] ratios, and its very large [4552]/[4542] ratio values, we conclude that B0 is the most probable spectral type for star #4. In the case of the B-type stars, it is not possible to use the criteria presented by Sota et al. (2011) to estimate luminosity classes. Instead, we used the results by Didelon (1982) for B-stars, and from the observed helium and hydrogen EW values we are able to assign luminosity class IV-V to this star.

## 4. DISCUSSION

A measure of the validity of the empirical methods presented in this paper for obtaining spectral classifications of O and B-type stars from APOGEE spectra is the degree with which such results compare with those from standard classification methods from blue optical spectra. Bear in mind, though, that the purpose of this work is not to propose a complete spectral type and luminosity classification system for O- and B-type stars based on APOGEE spectra, but rather, to show that the use of the APOGEE spectrograms can indeed produce reliable results. For stars suffering from high dust extinction, such as regions of very recent star formation, the proposed scheme should prove extremely valuable for identifying the OB stars, and their subsequent spectral classification. Indeed, current near-IR surveys like the VVV (VISTA Variables in the Via Lacteal survey - Minniti et al. 2010; Barbá et al. 2015) are finding several clusters with large numbers of candidate O and B stars photometrically throughout the Southern Galactic plane, and these are prime targets for spectroscopic confirmation with our APOGEE classification pipeline developed here.

In the case of the four selected stars for which we determined spectral types and luminosity classes independently from APOGEE and from blue optical spectra, the comparison yielded mixed results. The spectral types obtained based on NIR He II lines and optical data (shown respectively in Tables 4 and 6), for stars APOGEE-OB-01 (O7.5-O8 (NIR) versus O8 (blue)), and APOGEE-OB-02 (O8.5-O9 (NIR) versus O8.5 (blue)) agree very well. In the cases of APOGEE-OB-03 and APOGEE-OB-04, we followed a different approach, which was to estimate spectral types based on lower an upper limits obtained by setting EW (He II)=0 into Equations 3 and 5, and using the corresponding EW (Br11+Br13) values into Equations 1 and 2. The results for these two stars are also given in Tables 4 an 6 and clearly show that the derived spectral type for each star agree reasonably well regardless of which method (NIR or blue) was applied. This confirms that APOGEE data also provides good spectral type estimates even in cases where the He II 7-12 and 7-13 lines are absent in the NIR spectrum.

In this sense (as mentioned briefly in Section 1), the spectral classification of hot massive stars at infrared wavelengths is advantageous in terms of classifying sources suffering from moderate to high dust extinction: Analyses of H-band APOGEE spectra are, thus, valid alternatives to those based on optical wavelengths, and this is particularly important for Galactic regions that are heavily affected by dust. This is thanks to the fact that the hydrogen Brackett series combined with the moderately intense He II lines at 1.572 and 1.692 $\mu$m constitute a reliable set of absorption features which are relatively simple to measure. Furthermore, these results confirm those of Lenorzer et al. (2004) who showed that He line ratios in the infrared are well correlated with those in the optical, and thus can be used for spectral classification purposes.

## 5. CONCLUSION

In this work we analysed DR14 APOGEE spectra of a sample of 92 known OB stars. We demonstrated that the hydrogen Bracket (Br11 and Br13) and He II 7-12 and 7-13 line transitions, lying within the APOGEE H-band window, are suitable and reliable tools to be used in the identification and spectral classification studies of large number of mid-O to early B candidate stars. In this sense, we are in the process of applying the new methods described in this work to identify and classify moderately and highly reddened OB stars hidden among the ordinary Galactic disk population observed in the APOGEE surveys. In a more focused effort, we are currently using the APOGEE-North spectrograph to conduct a large survey ( 14 sq. deg.) in the direction of the massive star forming complexes W3, W4 and W5.

The authors would like to thank the anonymous referee by her/his useful comments and suggestions that contributed to improve the manuscript. A. Roman-Lopes acknowledges financial support provided in Chile by Comisión Nacional de Investigación Científica y Tecnológica (CONICYT) through the FONDECYT project 1170476 and by the QUIMAL project 130001. CR-Z and MT acknowledge support from UNAM-DGAPA-PAPIIT grants IN-108117 and IN-104316 Mexico. D.A.G.H. was funded by the Ramón y Cajal fellowship number RYC-2013-14182. D.A.G.H. and O.Z. acknowledge support provided by the Spanish Ministry of Economy and Competitiveness (MINECO) under grant AYA-2014-58082-P. JB and DM acknowledge support by the Ministry for the Economy, Development, and Tourism's "Programa Iniciativa Científica Milenio" through grant IC12009, awarded to The Millennium Institute of Astrophysics (MAS). We thank the staff of the Observatorio Astronómico Nacional in San Pedro Mártir for kind support during observations. Funding for the Sloan Digital Sky Survey IV has been provided by the Alfred P. Sloan Foundation, the U.S. Department of Energy Office of Science, and the Participating Institutions. SDSS-IV acknowledges support and resources from the Center for High-Performance Computing at the University of Utah. The SDSS web site is www.sdss.org. SDSS-IV is managed by the Astrophysical Research Consortium for the Participating Institutions of the SDSS Collaboration including the Brazilian Participation Group, the Carnegie Institution for Science, Carnegie Mellon University, the Chilean Participation Group, the French Participation Group, Harvard-Smithsonian Center for Astrophysics, Instituto de Astrofísica de Canarias, The Johns Hopkins University, Kavli Institute for the Physics and Mathematics of the Universe (IPMU) / University of Tokyo, Lawrence Berkeley National Laboratory, Leibniz Institut für Astrophysik Potsdam (AIP), Max-Planck-Institut für Astronomie (MPIA Heidelberg), Max-Planck-Institut für Astrophysik (MPA Garching), Max-Planck-Institut für Extraterrestrische Physik (MPE), National Astronomical Observatories of China, New Mexico State University, New York University, University of Notre Dame, Observatário Nacional / MCTI, The Ohio State University, Pennsylvania State University, Shanghai Astronomical Observatory, United Kingdom Participation Group, Universidad Nacional Autónoma de México, University of Arizona, University of Colorado Boulder, University of Oxford, University of Portsmouth, University of Utah, University of Virginia, University of Washington, University of Wisconsin, Vanderbilt University, and Yale University. This research has made use of the SIMBAD database, operated at CDS, Strasbourg, France - 2000, A&AS,143,9. This research has made use of the VizieR catalogue access tool, CDS, Strasbourg, France. The original description of the VizieR service was published in A&AS 143, 23. This research has made use of the services of the ESO Science Archive Facility. Based on observations collected at the European Organisation for Astronomical Research in the Southern Hemisphere under ESO programme 70.D-0110(A).

*Facilities:* Sloan Telescope, APOGEE, OANSPM:2.1m

*Software:* IRAF